\documentclass[journal]{IEEEtran}
\usepackage{graphicx}
\usepackage{cite}
\usepackage{picinpar}
\usepackage{amsmath}
\usepackage{url}
\usepackage{flushend}
\usepackage[latin1]{inputenc}
\usepackage{colortbl}
\usepackage{soul}
\usepackage{multirow}
\usepackage{pifont}
\usepackage{color}
\usepackage{alltt}
\usepackage[hidelinks]{hyperref}
\usepackage{enumerate}
\usepackage{breakurl}
\usepackage{epstopdf}
\usepackage{pbox}

\usepackage{fancyhdr}
\usepackage{lineno} 
\usepackage{bm}
\newtheorem{theorem}{Theorem}

\usepackage{amssymb}
\usepackage{stfloats}
\usepackage{algorithmic}
\usepackage[ruled,vlined]{algorithm2e}

\usepackage[table]{xcolor}
\usepackage{comment}
\usepackage[nottoc]{tocbibind}

\newcommand{\rv}[1]{{#1}}

\begin{document}

\title{Blockchain-Empowered Socially Optimal Transactive Energy System: Framework and Implementation 
}

\author{Qing~Yang,~\IEEEmembership{Member,~IEEE,}
        Hao~Wang,~\IEEEmembership{Member,~IEEE}
\thanks{This work is in part supported by the National Natural Science Foundation of China (project 61901280) and the FIT Academic Staff Funding of Monash University.}
\thanks{Q. Yang is with the Blockchain Technology Research Center (BTRC) and the College of Electronics and Information Engineering (CEIE), Shenzhen University, Shenzhen, Guangdong Province, PRC, e-mail: {yang.qing@szu.edu.cn}.}
\thanks{H. Wang is with Department of Data Science and Artificial Intelligence, Faculty of Information Technology, Monash University, Melbourne, VIC 3800, Australia, and Stanford Sustainable Systems Lab, Stanford University, CA 94305, USA, e-mail: hao.wang2@monash.edu.}
\thanks{Corresponding author: Hao Wang.}
}

\maketitle

\begin{abstract}
Transactive energy plays a key role in the operation and energy management of future power systems. However, the conventional operational mechanism, which follows a centralized design, is often less secure, vulnerable to malicious behaviors, and suffers from privacy leakage. In this work, we introduce blockchain technology in transactive energy to address these challenges. Specifically, we develop a novel blockchain-based transactive energy framework for prosumers and design a decentralized energy trading algorithm that matches the operation of the underlying blockchain system. We prove that the trading algorithm improves the individual benefit and guarantees the socially optimal performance, and thus incentivizes prosumers to join the transactive energy platform. Moreover, we evaluate the feasibility of the transactive energy platform throughout the implementation of a small-scale network of Internet of Things (IoT) devices and extensive simulations using real-world data. Our results show that this blockchain-based transactive energy platform is feasible in practice, and the decentralized trading algorithm reduces the user's individual cost by up to 77\% and lowers the overall cost by 24\%.
\end{abstract}


\markboth{IEEE TRANSACTIONS ON INDUSTRIAL INFORMATICS}%
{}

\SetAlFnt{\small}
\SetAlCapFnt{\small}

\section{Introduction}\label{sec:intro}
\IEEEPARstart{O}{ur} homes are going greener and smarter with the ubiquitous deployment of smart meters, renewable energy, and behind-the-meter energy storage. Smart homes \cite{tushar2014three}, equipped with renewable generators and energy storage, can smartly manage their household generation and demand. To further improve the efficiency and resiliency of the power system, smart homes are encouraged to exchange energy with each other; therefore, transactive energy over the smart grid attracts intense interest from both academia and industry.

As a feasible implementation of transactive energy, peer-to-peer (P2P) energy trading \cite{paudel2018peer} provides a prospective solution to benefit both the system and customers. With P2P energy trading, customers can exchange surplus energy via existing grids with others to gain benefits and improve the social welfare of the system. However, the following challenges must be addressed before P2P energy trading can be widely accepted and deployed in smart grids. First, the centralized energy market architecture, which is controlled by a central node (usually the grid operator), is prone to single-point failures and limits the system's trust. In such a system, the trading process is opaque to the users; thus, the users may not trust the matching result and price, which hinders users from participating in the energy trading system. Second, balancing the social welfare and individual user's benefit is difficult for P2P energy trading among independent users. Third, users' privacy (e.g., identity and energy consumption record) is vulnerable in centralized energy trading systems.

Recently, the booming of cryptocurrency and decentralized applications (DApps) facilitate the versatile application of the blockchain technology \cite{swan2015blockchain} in various areas. As the underlying technology of the cryptocurrency, blockchain is a decentralized ledger (or database) maintained by a group of nodes with the \textit{consensus protocol}, thus removing the need for an authoritative central node \cite{dinh2018untangling}. Furthermore, by supporting \textit{smart contracts}, blockchain (e.g., Ethereum\cite{eth}) provides a trustable computing platform, on which users can run generic computer programs. Regarding the social welfare and privacy of P2P energy trading, in \cite{wang2016incentivizing}, the authors developed an incentive mechanism that minimizes social costs and a distributed privacy-preserving energy trading algorithm for interconnected microgrids. Inspired by these ideas, we aim to develop an efficient and privacy-preserving transactive energy system.

In this work, to address the aforementioned challenges, we integrate blockchain with a novel decentralized P2P energy trading algorithm. The proposed trading algorithm consists of two levels: the user-level algorithm is run on user nodes for local cost optimization;
the blockchain-level algorithm is run on blockchain for system-wise energy trading optimization. The decentralized trading algorithm converges to the optimal trading outputs in a few rounds of interaction on the blockchain. The main contributions of this work are as follows.
\begin{enumerate}[1)]
  \item We develop and implement a trustable decentralized transactive energy platform based on blockchain technology that outputs correct, intact, and verifiable energy trading results.  
  \item We design an energy trading mechanism that optimizes social welfare as well as users' individual benefit.
  \item We develop a decentralized P2P energy trading algorithm consisting of two levels on the blockchain and user sides, preserving users' privacy. 
\end{enumerate}

The remainder of the paper is organized as follows. Section~\ref{sec:related} introduces the background and related works. Section~\ref{sec:model} describes the system model of the blockchain-based decentralized transactive energy platform. Section~\ref{sec:formulation} formulates the energy trading problem. Section~\ref{sec:solution} elaborates the decentralized P2P energy trading algorithm. Section~\ref{sec:eval} evaluates the proposed system with extensive experiments and simulations. Section~\ref{sec:conclusion} concludes this paper.

\section{Background and Related Works}\label{sec:related}
We first summarize the related works on energy trading in the context of the smart grid and Internet of Things (IoT). Then we discuss and compare the existing blockchain-empowered energy trading algorithms and applications.

Blockchain is a decentralized ledger (or database) maintained by all the nodes using a mechanism named consensus protocol \cite{swan2015blockchain}. A good tutorial and analysis on various consensus protocols can be found in \cite{natoli2019deconstructing}. Upon the shared ledger, the latest blockchain systems support smart contracts, which are a generic computer program developed by the users, e.g., Solidity on Ethereum \cite{eth}. In this paper, we use blockchain as a trustable and decentralized computing platform to implement our P2P energy trading algorithm to guarantee the transparency and efficiency of the trading process.

Recently, blockchain-based energy trading attracts much research interest. The possible applications of blockchain in networked microgrids was discussed in \cite{li2019blockchain}. \rv{Reference \cite{sabounchi2017towards} proposed an auction-based P2P energy trading algorithm and tested the algorithm on the Ethereum blockchain. In \cite{wang2019energy}, a crowdsourced energy system with P2P energy trading was implemented based on the Hyperledger blockchain. However, both \cite{sabounchi2017towards} and  \cite{wang2019energy} required the users to reveal all their power usage information including private information.} In \cite{li2017consortium}, the authors applied the consortium blockchain technology in the industrial internet of things (IIoT) to achieve secure energy trading and further introduced a credit-based payment scheme to support trading. \rv{However, \cite{li2017consortium} employed a centrally coordinated energy trading algorithm that exposes users' private information (e.g., users' power scheduling and battery parameters). To fill this gap, our work adopts a decentralized energy trading algorithm that does not reveal private information about the users and optimizes the overall social welfare.} LO3 Energy \cite{exergy} deployed a blockchain-based P2P energy trading platform named Exergy in the Brooklyn microgrid to facilitate online payments \cite{mengelkamp2018designing}. \rv{Exergy employed the blockchain technology only as a convenient payment tool for the users, but did not improve the efficiency of the trading system. Unlike \cite{exergy}, in order to implement a more efficient decentralized energy trading algorithm, our work utilizes blockchain not only as a payment tool but also as an energy trading and management platform.} In \cite{eric2017}, the authors proposed a decentralized algorithm for social welfare maximization and proved the feasibility of this algorithm using Ethereum-based simulations. Compared with \cite{eric2017}, our work targets on incentivizing individual users while maximizing the social welfare, and we tailor the blockchain design for IoT devices such as smart meters.

Privacy is a critical concern in the blockchain-based P2P energy trading system because all the trading information stored on the blockchain is public and  transparent, which is prone to be exploited by malicious attackers \cite{pnnl}. \rv{Reference \cite{aitzhan2016security} proposed to use multi-signature and asymmetric encryption to conceal users' private information and implemented this method on the blockchain.} In \cite{gai2019privacy}, the authors used a one-to-multiple account mapping algorithm to hide users' trading information. Reference \cite{li2017consortium} introduced a private account-pool at the aggregator to conceal the real identity of users. \rv{However, both \cite{li2017consortium} and \cite{gai2019privacy} adopted a centralized identity management method relying on a trustable central node to manage all the privacy information, so this method is vulnerable to the single-point failure of the central node.} A privacy model for the blockchain-based smart factory was introduced, and a white-list mechanism was proposed to improve the security in \cite{wan2019blockchain}. \rv{However, the method proposed in \cite{wan2019blockchain} depends on a private blockchain, in which only permitted users can join. In this work, we design a decentralized optimization algorithm that only discloses the users' energy-trading requests and works on the blockchain system.}

\section{System Model}\label{sec:model}
This section describes the model of the blockchain-empowered transactive energy system. The system consists of three layers: 1) the \textit{smart grid system layer} manages the smart grid components; 2) the \textit{energy trading layer} implements a decentralized P2P energy trading algorithm;
3) the \textit{blockchain layer} acts as a decentralized and trustable energy trading platform and a secure data communication network.

\begin{figure}[!t]
    \centering
    \includegraphics[width=8.7cm]{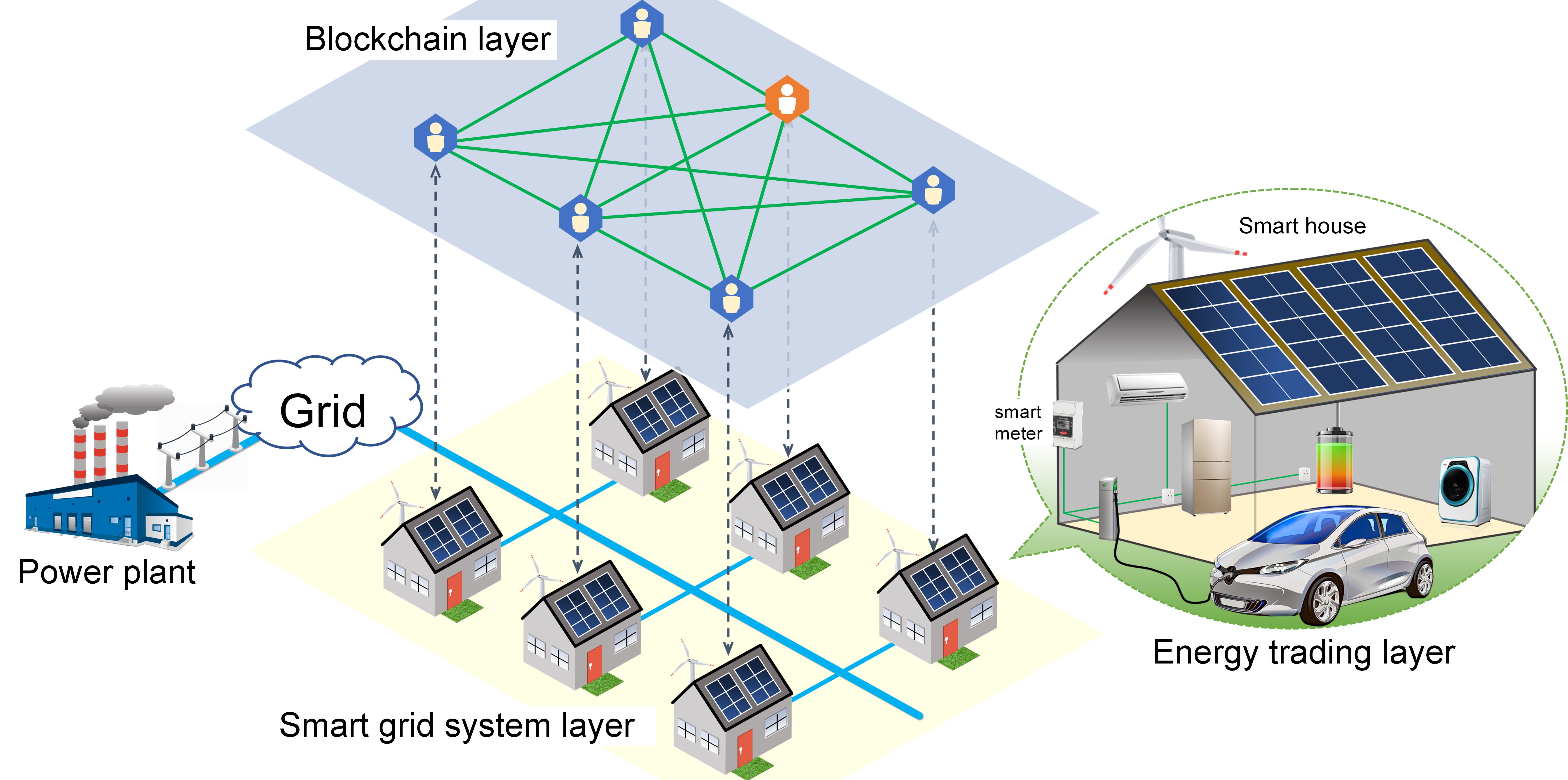}
    \vspace*{-3mm}
    \caption{The system model of the blockchain-empowered decentralized transactive energy platform.}
    \vspace*{-4mm}
    \label{f1:sysmod}
\end{figure}

\vspace*{-2mm}
\subsection{The Smart Grid System Layer}
 The main components in the smart grid layer include both supply from the grid and residential users who have local renewable generations and loads. We denote the set of residential users by $\mathcal{N} {=} \{1,..., N \}$, who use energy management systems to control their appliances as shown in Fig.~\ref{f1:sysmod}. Users also have small-scale renewable generators (including rooftop PV panels and wind turbines), energy storage units (like Lithium-ion batteries), and bidirectional DC-AC inverters. Thus, users can trade energy with others over an operational horizon $\mathcal{H} {=} \{1,..., H \}$ with $H$ time slots. Moreover, the load profiles of users exhibit great diversity \cite{xu2019novel}, making energy exchange among users and their neighbors a promising solution. We present the model of users and the P2P energy trading platform in the following. 

\subsubsection{User Model}
The user model consists of the load model, the electricity supply model, and the battery storage model.
User $i$'s home appliances can be classified into three types. The first type of appliance load includes heating, ventilation, and air conditioning (HVAC) unit that often consumes a large amount of energy and is adjustable to provide flexibility in power scheduling \cite{yang2020cooperative}. The second type of appliance load is shiftable over time; for example, the use of washing machines and dryers is flexible in time according to the users' schedule. The rest of the load is treated as inflexible load as it is used to meet the basic needs, e.g., lighting, refrigerator, and cooking.

The HVAC of user $i$ consumes power $p_i^{\text{AC}}[t]$ to control the indoor temperature $T_i^{\text{IN}}[t]$ against the outdoor temperature $T_i^{\text{OUT}}[t]$ in time slot $t$. Let $p_i^{\text{AC}}[t]$ denote the electricity consumption of the heating or cooling process incurred by the HVAC system. Specifically, the dynamics of the indoor temperature \cite{cui2019} is modeled as
    \begin{align}
            T_i^{\text{IN}}[t] = T_i^{\text{IN}}[t-1] - & \frac{1}{C_i R_i} ( T_i^{\text{IN}}[t-1] - T_i^{\text{OUT}}[t] \nonumber\\
            & + \eta_i R_i p_i^{\text{AC}}[t] ), ~ \forall i \in \mathcal{N}, t \in \mathcal{H}, \label{constraint-load1}
    \end{align}
where $C_i$ and $R_i$ are the HVAC system parameters which indicates its working efficiency; $\eta_i$ indicates the HVAC's working modes, specifically, a positive $\eta_i$ for cooling and a negative $\eta_i$ for heating.

The indoor temperature is usually stabilized at a user-preferred value  $T_i^{\text{REF}}$ by the HVAC, and any deviation will discomfort the user. We model the discomfort as a cost function of user $i$'s experience: 
    \begin{equation}
            C_i^{\text{AC}} = \beta_i^{\text{AC}} \sum\nolimits_{t \in \mathcal{H}} \left( T_i^{\text{IN}}[t] - T_i^{\text{REF}} \right)^{2}, ~ \forall i \in \mathcal{N}, \label{objective-load1}
    \end{equation}
which measures the discomfort that users feel if the indoor temperature deviates. Also, the indoor temperature should be controlled within a range that users can tolerate, hence 
    \begin{equation}
            \underline{T}_i^{\text{IN}} \leq  T_i^{\text{IN}}[t] \leq  \overline{T}_i^{\text{IN}}, ~ \forall i \in \mathcal{N}, t \in \mathcal{H}, \label{constraint-load2}
    \end{equation}
where $\underline{T}_i^{\text{IN}}$ and $\overline{T}_i^{\text{IN}}$ are the lowerbound and upperbound of the tolerable indoor temperature of user $i$.

The flexible load $p_i^{\text{F}}[t]$ of user $i$ in time slot $t$ can be shifted over time, but the flexible load scheduling for user $i$ should satisfy the following constraints:
    \begin{align}
            \sum\nolimits_{t \in \mathcal{H}_i} p_i^{\text{F}}[t] = D_i^{\text{F}}, &~ \forall i \in \mathcal{N}, \label{constraint-load3} \\
            \underline{P}_i^{\text{F}}[t] \leq  p_i^{\text{F}}[t] \leq  \overline{P}_i^{\text{F}}[t], &~ \forall i \in \mathcal{N}, t \in \mathcal{H}_i, \label{constraint-load4}
    \end{align}
in which $\mathcal{H}_i \subseteq \mathcal{H}$ denotes the time window of user $i$ for scheduling the flexible load. 
Constraint \eqref{constraint-load3} specifies the total flexible load $D_i^{\text{F}}$ of user $i$ within the time window $\mathcal{H}_i$. The flexible load is also bounded within $ \underline{P}_i^{\text{F}}[t]$ and $\overline{P}_i^{\text{F}}[t]$. 

Though the flexible load can be shifted, user $i$ often has a preferred schedule $P_i^{\text{REF}}[t]$ reflecting the most comfortable schedule. When the actual load $p_i^{\text{F}}[t]$ deviates from the preferred schedule $P_i^{\text{REF}}[t]$, user $i$ will suffer discomfort as
    \begin{align}
            C_i^{\text{F}} = \beta_{i}^{\text{F}} \sum\nolimits_{t \in \mathcal{H}_i} \left( p_i^{\text{F}}[t] - P_i^{\text{REF}}[t] \right)^{2}, \label{objective-load2}
    \end{align}
which penalizes the flexible load deviation. The coefficient $\beta_{i}^{\text{F}}$ denotes the discomfort cost, indicating the sensitivity of user $i$ towards the deviation from the preferred power consumption.

The inflexible load $P_i^{\text{IL}}[t]$ in time slot $t$ corresponds to the energy consumption of appliances that cannot be shifted, e.g., lighting and refrigerators. Different from the HVAC load $p_i^{\text{AC}}[t]$ and flexible load $p_i^{\text{F}}[t]$, user $i$ does not schedule $P_i^{\text{IL}}[t]$.

\subsubsection{Supply Model}
Users can purchase electricity from the grid denoted as $p_i^{\text{G}}[t]$ and use renewable energy locally denoted as $p_i^{\text{RE}}[t]$. We will discuss the trading of electricity with other users later. The grid power purchase and renewable power supply should satisfy the following constraints:
    \begin{align}
            0 \leq  p_i^{\text{G}}[t] \leq P_i^{\text{G}}, &~ \forall i \in \mathcal{N}, t \in \mathcal{H}, \label{constraint-load6}\\
            0 \leq p_i^{\text{RE}}[t] \leq P_i^{\text{RE}}[t], &~ \forall i \in \mathcal{N}, t \in \mathcal{H}, \label{constraint-load5} 
    \end{align}
in which $p_i^{\text{G}}[t]$ and $p_i^{\text{RE}}[t]$ are non-negative and upper-bounded by the power line capacity $P_i^{\text{G}}$ and available renewable generation $P_i^{\text{RE}}[t]$, respectively.

For the electricity bill to the grid, user $i$ needs to pay a two-part tariff:
    \begin{align}
            C_i^{\text{G}} = \pi_1^{\text{G}} \sum\nolimits_{t\in \mathcal{H}} p_i^{\text{G}}[t] + \pi_2^{\text{G}} \max_{t\in \mathcal{H}} p_i^{\text{G}}[t], \label{objective-supply} 
    \end{align}
which consists of an energy charge $\pi_1^{\text{G}} \sum_{t\in \mathcal{H}} p_i^{\text{G}}[t]$ and a peak charge $\pi_2^{\text{G}} \max_{t\in \mathcal{H}} p_i^{\text{G}}[t]$, where $\pi_1^{\text{G}}$ and $\pi_2^{\text{G}}$ denote the energy price and peak price, respectively. In addition to the energy charge that counts the total energy consumption, the peak charge incentivizes peak shaving from users.

\subsubsection{Battery Model} 
Each user has a battery unit to charge harvested renewable energy or grid power, and to discharge later to serve the load. The battery capacity of user $i$ is denoted by $E_i^{\text{B}}$; the charging power and discharging power are denoted by $p_i^{\text{CH}}[t]$ and $p_i^{\text{DIS}}[t]$, respectively. The energy storage level of the battery in time slot $t$ is denoted by $e_i^{\text{B}}[t]$ and the battery operation should satisfy the following constraints:
    \begin{align}
            e_i^{\text{B}}[t] {=} e_i^{\text{B}}[t{-}1] {+} \eta_i^{\text{CH}} p_i^{\text{CH}}[t] {-} \frac{1}{\eta_i^{\text{DIS}}} p_i^{\text{DIS}}[t], \forall i {\in} \mathcal{N}, t {\in} \mathcal{H}, \label{constraint-load7} \\
            \underline{\alpha}_i^{\text{B}} E_i^{\text{B}} \leq e_i^{\text{B}}[t] \leq \overline{\alpha}_i^{\text{B}} E_i^{\text{B}}, ~ \forall i \in \mathcal{N}, t \in \mathcal{H}, \label{constraint-load8} \\
            0 \leq  p_i^{\text{CH}}[t] \leq P_i^{\text{CH}}, ~ \forall i \in \mathcal{N}, t \in \mathcal{H}, \label{constraint-load9} \\
            0 \leq  p_i^{\text{DIS}}[t] \leq P_i^{\text{DIS}}, ~ \forall i \in \mathcal{N}, t \in \mathcal{H}, \label{constraint-load10}
    \end{align}
where $\eta_i^{\text{CH}} \in [0,1]$ and $\eta_i^{\text{DIS}} \in (0,1]$ denote the charging and discharging efficiency of the battery, respectively. Constraint \eqref{constraint-load7} specifies the dynamics of the storage level with respect to the charging power and discharging power. The operational range of the energy level is bounded within $\underline{\alpha}_i^{\text{B}} E_i^{\text{B}} $ and $\overline{\alpha}_i^{\text{B}} E_i^{\text{B}}$, in which $\underline{\alpha}_i^{\text{B}} \in [0,1)$ and $\overline{\alpha}_i^{\text{B}} \in (0,1]$ denote the lower-bound and upper-bound fractions of the battery capacity. 

In addition, the charging power $p_i^{\text{CH}}[t]$ and discharging power $p_i^{\text{DIS}}[t]$ are both non-negative and upper-bounded by the rated charge and discharge power $P_i^{\text{CH}}$ and $P_i^{\text{DIS}}$, respectively. The battery operation also incurs degradation cost as
    \begin{align}
            C_i^{\text{B}} = \beta_i^{\text{B}} \sum\nolimits_{t\in \mathcal{H}} \left( p_i^{\text{CH}}[t] + p_i^{\text{DIS}}[t] \right), \label{objective-battery}
    \end{align}
where $\beta_i^{\text{B}}$ is the operational cost coefficient of the battery. 

\subsection{The Energy Trading Layer}
\rv{This layer focuses on the modeling of the energy exchange and payments among users. Users can exploit the diversities of supply and demand profiles \cite{wang2014hybrid} and trade energy with each other for mutual benefits. The energy trading layer enables active energy exchange among neighbor households without changing the existing smart grid infrastructure (as described in the smart grid layer). This is a great advantage of the developed energy trading, which does not demand a full upgrade of infrastructure but requires smart trading decision-making algorithms and information exchange platform. }

In the P2P energy trading system shown in Fig.~\ref{f1:sysmod}, user $i$ can form trading pairs with user $j \in \mathcal{N} \backslash i$ to exchange energy. Note that $p_{i,j}^{\text{ET}}[t] >0$ if user $i$ purchases energy from user $j$ in time slot $t$; otherwise, $p_{i,j}^{\text{ET}}[t] <0$ if user $i$ sells energy to user $j$. Over the energy exchange horizon $\mathcal{H}$, user $i$ also determines the associated payment $\pi_{i,j}^{\text{ET}}$ to user $j$. Similarly, $\pi_{i,j}^{\text{ET}}$ is positive if user $i$ makes payment to user $j$, negative if user $i$ receives payment from user $j$. 

The blockchain-enabled transactive energy platform provides a free and open marketplace for residential users to trade energy and allow participated users to earn profits. \rv{During the trading process, the users' trading decisions $p_{i,j}^{\text{ET}}[t]$ and payment decisions $\pi_{i,j}^{\text{ET}}$ are transmitted to the underlying blockchain as transactions. The decentralized energy trading algorithm uses these transactions to compute the optimal energy trading and payment results. Then the energy trading among users is executed on the distribution network, and the payment is settled using the token of the blockchain.} 

Since the users are located close to each other, we assume that the loss of energy during the exchange is negligible. Therefore, we have the following clearing constraints for the P2P energy trading and associated payment process:
    \begin{align}
        p_{i,j}^{\text{ET}}[t] + p_{j,i}^{\text{ET}}[t] = 0,&~\forall t \in \mathcal{H},~\forall i \in \mathcal{N},~\forall j \in \mathcal{N} \backslash i, \label{constraint-trading1} \\
        \pi_{i,j}^{\text{ET}} + \pi_{j,i}^{\text{ET}} = 0,&~\forall i \in \mathcal{N},~\forall j \in \mathcal{N} \backslash i, \label{constraint-trading2} 
    \end{align}
where constraint \eqref{constraint-trading1} clears the energy trading between each trading pair of users $i$ and $j$ in time slot $t$, and constraint \eqref{constraint-trading2} clears the payment between the trading pair of users $i$ and $j$. We next present how the users optimize the energy exchange amount $p_{i,j}^{\text{ET}}[t]$ and payment $\pi_{i,j}^{\text{ET}}$.

\rv{The energy trading layer focuses on the algorithmic design of the decentralized P2P energy trading. In this layer, we modeled the P2P energy trading problem in Section \ref{subsec:trading} and developed a decentralized solution in Section \ref{sec:solution}. 
To facilitate energy trading, we need not only energy-trading decision making but also support from a trustable information-exchange platform. Therefore, we develop a separate blockchain layer from the energy trading layer. The blockchain layer focuses on the implementation of the underlying blockchain system to support the energy trading layer. We will consider the practical issues of deploying a feasible blockchain system on the IoT devices (e.g., the smart meters) that have limited hardware resources and network bandwidth. The detailed design of the blockchain system will be introduced in Section \ref{sec:blockchainlayer}.}

\subsection{The Blockchain Layer}\label{sec:blockchainlayer}

\begin{figure}[!t]
    \centering
    \includegraphics[width=8.7cm]{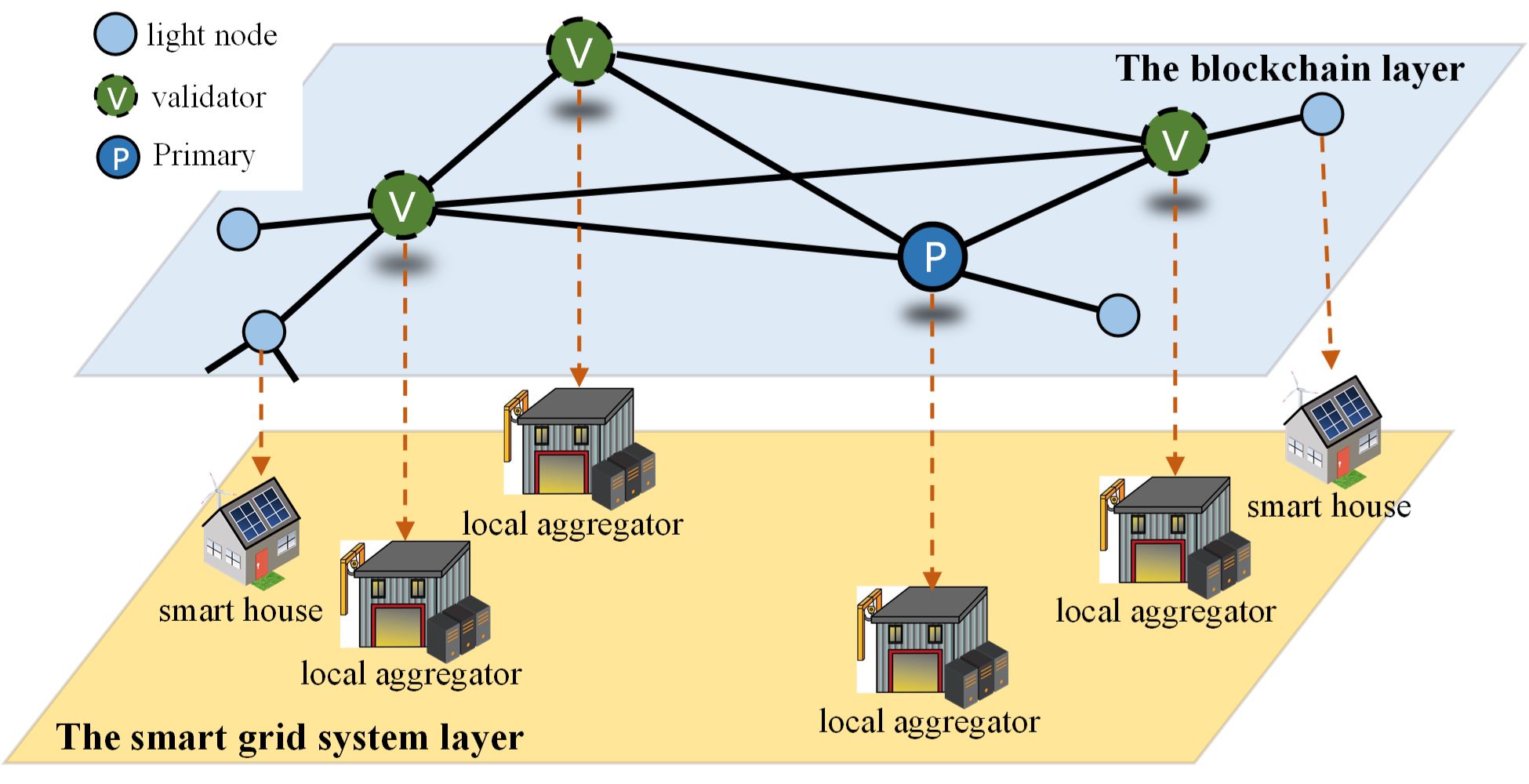}
    \caption{\rv{The system model of the blockchain network and the mapping between the smart grid system and the blockchain system.}}
    \vspace*{-3mm}
    \label{fn:mapping}
\end{figure}

\subsubsection{Data Structure of the Blockchain System}
To enable transactive energy over the blockchain, we tailor the underlying blockchain system for the IoT devices (e.g., smart meters) 
that prevails in smart grid and industrial IoT environments. In this work, we adopt the conventional chain-based structure 
for our blockchain instead of the DAG-based structure, which is popular in IoT systems \cite{conflux,huang2019towards}, to achieve faster transaction confirmation and support smart contracts. \rv{The blockchain nodes interact with the blockchain for energy trading and payment by emitting \textit{transactions}. All the transactions emitted during the current block interval are packed into one block during one round of the consensus protocol.}

\subsubsection{Trading and Payment Transactions}\label{ss:account}
The blockchain provides a built-in token system to ease the payment in energy trading. Unlike \cite{li2017consortium} that encourages nodes to ``mine'' for more tokens, we pre-allocate tokens and peg the tokens to fiat money such as US dollars. This mechanism stabilizes the value of the token to avoid speculation and protect token-holders' interests.

To support energy trading, we define two types of transactions: 1) the token transfer transaction allows a node to transfer token to others for the payment; 2) the energy trading transaction, which contains the user's trading information $p_{i}^{\mathrm{ET}}(t)$ and $\pi_{i}^{\mathrm{ET}}(t)$, exchanges trading information with the smart contract for energy trading. We implement the decentralized P2P energy trading algorithm in a smart contract and deploy it on the blockchain. Residential users call this smart contract to trade energy by emitting a transaction on the blockchain. With several rounds of interaction, the smart contract outputs the optimal trading schedule, then all users execute the trading schedule and pay with the token. 

\subsubsection{Consensus Protocol}\label{ss:consensus}
To support secure and efficient P2P energy trading on the blockchain, we design a practical Byzantine-fault-tolerance (PBFT) style consensus protocol \cite{castro1999practical} to synchronize the ledger states of all the nodes. We classify all nodes into two categories: light nodes and validator nodes, as illustrated in Fig.~\ref{fn:mapping}. \textit{Light nodes} can trade electricity with other nodes and commit payments by submitting transactions and calling smart contracts, but do not participate in the consensus process. \rv{In the smart grid, users who access the blockchain network with smart meters act as light nodes.} \textit{Validator nodes} participate in the consensus process and verify the blocks. Among the validators, one is chosen as the \textit{primary validator} who is responsible for collecting transactions and packing them into blocks. Being a validator requires higher computational power and network bandwidth, so we suggest the local aggregators or residential users who have competent hardware be the validators. \rv{The primary validator is chosen in a round-robin manner with pre-schedule order. Specifically, if the current primary validator successfully packages a block or failed to reach consensus in the prescribed time, then the next validator will automatically become the new primary validator \cite{quorum}. To incentivize the validators, the primary validator will receive the block rewards and transaction fees, paid in the blockchain tokens, if it successfully packages a block.}

The trust and security of the blockchain system are guaranteed by the PBFT consensus protocol and digital signature algorithm. Here, ``trust'' means that nobody can tamper the data recorded on the blockchain or manipulate the results of the energy trading algorithm, which is implemented as smart contract. The validators use the elliptic curve digital signature algorithm (ECDSA) to sign all the messages transmitted during the consensus process. The digital signature authenticates the identity of the validator and also guarantees the integrity of the transmitted message, thus builds trust among the validators.

\rv{We consider two attack vectors that can undermine the blockchain. First, a malicious validator may try to tamper the results of the energy trading for its own interest. However, the processing and results of the decentralized energy trading algorithm, which is implemented in smart contract, are audited and approved by all the validators during the PBFT consensus. As shown in \cite{castro1999practical}, the PBFT consensus protocol can tolerate this error if the number of malicious validators is less than one-third of the total validators. Therefore, this attack cannot succeed unless more than one-third of the validators collude to act maliciously, which rarely happens. Second, a malicious primary validator may try to impede the consensus process by sending conflicting messages to other validators. To counter this attack, each validator has a timer to measure the time that the current primary validator consumes on consensus. If the consensus pauses for a period of time that is longer than the timeout, the timer will trigger a view-change and the next validator will become the primary validator to continue the consensus process.}


\section{Problem Formulation}\label{sec:formulation}
Following the presentation of the system models, in this section, we consider two operational scenarios, i.e., without and with energy trading among users. In the standalone scenario, all the users manage their own energy supply and consumption independently so there is no energy exchange among the users. In the energy-trading scenario, users are allowed to trade energy with other users via the blockchain-enabled transactive energy platform. We will take the standalone scenario as the baseline for comparison.
\vspace*{-4mm}
\subsection{Standalone Scenario (SA): Without Energy Trading}\label{sec:standalone}
We first consider the scenario reflecting the present practice in which the users do not trade energy with each other. The standalone users schedule their energy supply, battery storage, and energy consumption of appliances to minimize the operating costs. 
For simplicity of notation, we use the short notation as follows: 
$\bm{p}_i^{\text{RE}} {=} \{ p_i^{\text{RE}}[t],\forall t {\in} \mathcal{H}\}$,
$\bm{p}_i^{\text{G}} {=} \{ p_i^{\text{G}}[t],\forall t {\in} \mathcal{H}\}$,
$\bm{p}_i^{\text{AC}} {=} \{ p_i^{\text{AC}}[t],\forall t {\in} \mathcal{H}\}$,
$\bm{p}_i^{\text{F}} {=} \{ p_i^{\text{F}}[t],\forall t {\in} \mathcal{H}\}$,
$\bm{p}_i^{\text{CH}} {=} \{ p_i^{\text{CH}}[t],\forall t {\in} \mathcal{H}\}$,
$\bm{p}_i^{\text{DIS}} {=} \{ p_i^{\text{DIS}}[t],\forall t {\in} \mathcal{H}\}$,
and 
$\bm{e}_i^{\text{B}} {=} \{ e_i^{\text{B}}[t], \forall t {\in} \mathcal{H}\}$.

To schedule the energy supply and consumption of the smart home, user $i$ needs to balance the total power supply and demand in every time slot.

Therefore,
    \begin{align}
            p_i^{\text{RE}}[t] + p_i^{\text{G}}[t] + p_i^{\text{DIS}}[t] = &p_i^{\text{AC}}[t] + p_i^{\text{F}}[t] + P_i^{\text{IL}}[t] \nonumber\\
             & + p_i^{\text{CH}}[t], ~ \forall i \in \mathcal{N}, t \in \mathcal{H}, \label{constraint-load11}
    \end{align}
where the left-hand side represents the total supply $p_i^{\text{RE}}[t] + p_i^{\text{G}}[t] + p_i^{\text{DIS}}[t]$, and the right-hand side denotes the total demand from the HVAC load $p_i^{\text{AC}}[t]$, the flexible load $p_i^{\text{F}}[t]$, the inflexible load $P_i^{\text{IL}}[t]$ and the battery charge $p_i^{\text{CH}}[t]$.

Based on the model in Section~\ref{sec:model}, the overall operating cost of user $i$ in the standalone scenario is
\begin{align}
            C_i^{\text{O}}(\bm{p}_i^{\text{G}}, \bm{p}_i^{\text{AC}}, \bm{p}_i^{\text{F}}, \bm{p}_i^{\text{CH}}, \bm{p}_i^{\text{DIS}}) \triangleq &
            C_i^{\text{G}}(\bm{p}_i^{\text{G}}) + C_i^{\text{AC}}(\bm{p}_i^{\text{AC}}) + C_i^{\text{F}}(\bm{p}_i^{\text{F}})  \nonumber \\
            & +  C_i^{\text{B}}(\bm{p}_i^{\text{CH}},\bm{p}_i^{\text{DIS}}), \label{objective-operatingcost}
\end{align}
where $C_i^{\text{G}}(\bm{p}_i^{\text{G}})$ denotes the electricity bill to the grid, $C_i^{\text{AC}}(\bm{p}_i^{\text{AC}})$ and  $C_i^{\text{F}}(\bm{p}_i^{\text{F}})$ denote the discomfort costs of scheduling the HVAC load and the flexible load, and $C_i^{\text{B}}(\bm{p}_i^{\text{CH}},\bm{p}_i^{\text{DIS}})$ denotes the battery operating cost.

User $i$ aims to minimize \eqref{objective-operatingcost}, and we formulate user $i$'s energy management problem $\textbf{EMP}_i$ as 

\textbf{EMP$_i$}: Energy Management Problem for User $i$
    \begin{equation*}
        \begin{aligned}
            &\text{minimize} && C_i^{\text{O}}(\bm{p}_i^{\text{G}}, \bm{p}_i^{\text{AC}}, \bm{p}_i^{\text{F}}, \bm{p}_i^{\text{CH}}, \bm{p}_i^{\text{DIS}}) \\
            &\text{subject to} && \text{\eqref{constraint-load1}}, \text{\eqref{constraint-load2} to \eqref{constraint-load4}},  \text{\eqref{constraint-load5} to \eqref{constraint-load6}}, \text{\eqref{constraint-load7} to \eqref{constraint-load10}},
            \text{\eqref{constraint-load11}}\\
            &\text{variables:} &&
            \bm{p}_i^{\text{RE}}, \bm{p}_i^{\text{G}}, \bm{p}_i^{\text{AC}}, \bm{p}_i^{\text{F}}, \bm{p}_i^{\text{CH}}, \bm{p}_i^{\text{DIS}}, \bm{e}_i^{\text{B}},
        \end{aligned}
    \end{equation*}
which is a convex optimization problem and can be solved by standard convex optimization techniques. We denote $\bar{C}_i^{\text{O}}$ as the minimized operating cost of \textbf{EMP$_i$} and use it as an upper-bound for the overall cost of user $i$ in the energy-trading scenario. We will explain it in details in Section \ref{subsec:trading}.

\vspace*{-3mm}
\subsection{Energy-Trading Scenario (ET): With Energy Trading}\label{subsec:trading}
Next, we consider an emerging scenario where users not only manage their energy supply and demand locally but also interact with other users to exchange energy. 
By trading energy over the blockchain-enabled platform, users can leverage the physical and cyber connectivity and the diversity of the users' energy profiles to achieve mutual benefits. 

For user $i$, the energy supply/demand balance constraint in the energy-trading scenario is
    \begin{equation}
        \begin{aligned}
            p_i^{\text{RE}}[t] & + p_i^{\text{G}}[t] + p_i^{\text{DIS}}[t] + \sum\nolimits_{j \in \mathcal{N} \backslash i} p_{i,j}^{\text{ET}}[t] = \\
            & p_i^{\text{AC}}[t] + p_i^{\text{F}}[t] + P_i^{\text{IL}}[t] + p_i^{\text{CH}}[t], ~ \forall i \in \mathcal{N}, t \in \mathcal{H}, \label{constraint-load12}
        \end{aligned}
    \end{equation}
where $\sum_{j \in \mathcal{N} \backslash i} p_{i,j}^{\text{ET}}[t]$ is the total energy traded between user $i$ and all other users $j \in \mathcal{N} \backslash i$ in time slot $t$. 

In addition to the operating cost $C_i^{\text{O}}(\bm{p}_i^{\text{G}}, \bm{p}_i^{\text{AC}}, \bm{p}_i^{\text{F}}, \bm{p}_i^{\text{CH}}, \bm{p}_i^{\text{DIS}})$, user $i$ also pays for energy trading. We denote the energy trading amount and payment of user $i$ by $\boldsymbol{p}_i^{\text{ET}} {=} \{ p_{i,j}^{\text{ET}}[t], \forall t \in \mathcal{H}, \forall j \in \mathcal{N} \backslash i \}$ and $\boldsymbol{\pi}_i^{\text{ET}} {=} \{ \pi_{i,j}^{\text{ET}}, \forall j \in \mathcal{N} \backslash i \}$, respectively. The energy-trading payment cost of user $i$ is
    \begin{align}
        C_{i}^{\text{ET}} (\boldsymbol{\pi}_i^{\text{ET}}) = \sum\nolimits_{j \in \mathcal{N} \backslash i} \pi_{i,j}^{\text{ET}}. \label{objective-tradingpayment}
    \end{align}
    
Users trade energy with others to gain some extra benefit due to the independent and selfish nature. Otherwise, if users cannot benefit from the energy trading, they can always refuse to trade energy and operate standalone to achieve the minimized costs $\bar{C}_i^{\text{O}}$ in the standalone scenario. To facilitate an effective energy trading platform, we need to ensure that all participated users are better off. In other words, users' costs should be no greater than the baseline costs when they do not trade energy, i.e., 
    \begin{equation}
        \begin{aligned}
            C_i^{\text{O}}(\bm{p}_i^{\text{G}}, \bm{p}_i^{\text{AC}}, \bm{p}_i^{\text{F}}, \bm{p}_i^{\text{CH}}, \bm{p}_i^{\text{DIS}}) {+} C_{i}^{\text{ET}} (\boldsymbol{\pi}_i^{\text{ET}})
            {\leq} \bar{C}_i^{\text{O}}, \forall i {\in} \mathcal{N}, \label{constraint-trading3}
        \end{aligned}
    \end{equation}
in which the overall cost of user $i$ consists of the operating cost $C_i^{\text{O}}(\bm{p}_i^{\text{G}}, \bm{p}_i^{\text{AC}}, \bm{p}_i^{\text{F}}, \bm{p}_i^{\text{CH}}, \bm{p}_i^{\text{DIS}})$ and trading payment $C_{i}^{\text{ET}} (\boldsymbol{\pi}_i^{\text{ET}})$. Note that the minimized operating cost $\bar{C}_i^{\text{O}}$ serves as a baseline or an outside choice for user $i$. 

Moreover, constraint \eqref{constraint-trading3} only guarantees that all users will not be worse off but does not specify how to allocate benefits to users. To ensure a fair allocation among all participated users, we formulate the energy exchange problem as a Nash bargaining problem targeting for proportional fairness. The energy trading optimization problem (ETOP) is formulated as 

    \textbf{ETOP}: Energy Trading Optimization Problem
    \begin{align*}
        &\text{maximize} && \prod_{i\in\mathcal{N}} \! \left[ \bar{C}_i^{\text{O}} {-} C_i^{\text{O}}(\bm{p}_i^{\text{G}}, \bm{p}_i^{\text{AC}}, \bm{p}_i^{\text{F}}, \bm{p}_i^{\text{CH}}, \bm{p}_i^{\text{DIS}}) {-} C_{i}^{\text{ET}} (\boldsymbol{\pi}_i^{\text{ET}}) \right] \\
        &\text{subject to} && 
        \text{\eqref{constraint-load1}},\text{\eqref{constraint-load2} to \eqref{constraint-load4}},  \text{\eqref{constraint-load5} to \eqref{constraint-load6}}, \text{\eqref{constraint-load7} to \eqref{constraint-load10}}, \text{\eqref{constraint-trading1} to \eqref{constraint-trading2}}, \\
            &&& \text{\eqref{constraint-load12}}, \text{\eqref{constraint-trading3}}\\
        &\text{variables:} &&
        \{ \bm{p}_i^{\text{RE}}, \bm{p}_i^{\text{G}}, \bm{p}_i^{\text{AC}}, \bm{p}_i^{\text{F}}, \bm{p}_i^{\text{CH}}, \bm{p}_i^{\text{DIS}}, \bm{e}_i^{\text{B}}, \bm{p}_i^{\text{ET}}, \bm{\pi}_i^{\text{ET}}, \forall i {\in} \mathcal{N} \},
    \end{align*}
where $\bar{C}_i^{\text{O}} - C_i^{\text{O}}(\bm{p}_i^{\text{G}}, \bm{p}_i^{\text{AC}}, \bm{p}_i^{\text{F}}, \bm{p}_i^{\text{CH}}, \bm{p}_i^{\text{DIS}}) - C_{i}^{\text{ET}} (\boldsymbol{\pi}_i^{\text{ET}})$ is the cost reduction of user $i$ comparing the standalone scenario with the energy-trading scenario. The objective function of Problem \textbf{ETOP} is the Nash product, which guarantees proportional fairness in allocating the trading benefits among users. Solving Problem \textbf{ETOP} usually requires a central node that collects all the users' information. However, the blockchain-enabled transactive energy platform is decentralized and thus we will present a decentralized algorithm design in Section~\ref{sec:solution}.

\section{Problem Analysis And Algorithm Design}\label{sec:solution}
In this section, we analyze the structure of the optimal solution to Problem \textbf{ETOP} and design a decentralized algorithm to solve Problem \textbf{ETOP}. In addition, we implement the algorithm in the blockchain-based transactive energy platform to analyze its feasibility and performance.
\vspace*{-3mm}
\subsection{Problem Analysis}
The solution to Problem \textbf{ETOP} includes energy scheduling decisions $\{\bm{p}_i^{\text{RE}}, \bm{p}_i^{\text{G}}, \bm{p}_i^{\text{AC}}, \bm{p}_i^{\text{F}}, \bm{p}_i^{\text{CH}}, \bm{p}_i^{\text{DIS}}, \bm{e}_i^{\text{B}}\}$, energy trading decisions $\{\bm{p}_i^{\text{ET}}\}$, and energy trading payment decisions $\{\bm{\pi}_i^{\text{ET}}\}$ of each user $i \in \mathcal{N}$. These decision variables are coupled across all users, hence it is unrealistic to collect all the users' decisions and solve Problem \textbf{ETOP} in a centralized manner, which would raise serious privacy concerns. Moreover, users are rational and aim to minimize their own costs while collaborating with each other to exchange energy in the platform. Yet, how such rational behaviors of users affect the overall performance of the trading platform or all the participated users as a whole remains unclear. Therefore, we first analyze the performance of the energy trading system and have proved the socially optimum as presented in Theorem~1.
    
    \begin{theorem}
        Problem \textbf{ETOP} minimizes the total cost of users $i \in \mathcal{N}$ who participate in the transactive energy platform.
    \end{theorem}
    
Theorem~1 shows that the rational behaviors of users who minimize their own costs in the energy exchange lead to the socially optimal performance for the platform, which is favored not only by the users but also the utility companies and policymakers. 
We omit the proof due to space limitation and will present the proof upon request. 

Per Theorem~1, Problem \textbf{ETOP} minimizes the total costs of the users; thus we utilize this property to decompose Problem \textbf{ETOP} into two subproblems. The first subproblem is the \textit{total cost minimization problem} (TCMP), which optimizes the energy scheduling and energy trading of all users to minimize the social operating cost. The second subproblem is the \textit{trading benefit allocation problem} (TBAP) that determines the trading payments of all users. Similar to the Problem \textbf{ETOP}, we define Problem \textbf{TCMP} in its mathematical form as
    
\textbf{TCMP}: Total Cost Minimization Problem
    \begin{equation*}
        \begin{aligned}
            &\text{minimize} {}&&{} \sum_{i\in\mathcal{N}} C_i^{\text{O}}(\bm{p}_i^{\text{G}}, \bm{p}_i^{\text{AC}}, \bm{p}_i^{\text{F}}, \bm{p}_i^{\text{CH}}, \bm{p}_i^{\text{DIS}}) \\
            &\text{subject to} {}&&{} 
            \text{\eqref{constraint-load1}}, \text{\eqref{constraint-load2} to \eqref{constraint-load4}},~\text{\eqref{constraint-load6} to \eqref{constraint-load5}}, \text{\eqref{constraint-load7} to \eqref{constraint-load10}}, \text{\eqref{constraint-trading1}}, \text{\eqref{constraint-load12}}\\
            &\text{variables:} {}&&{}
            \{ \bm{p}_i^{\text{RE}}, \bm{p}_i^{\text{G}}, \bm{p}_i^{\text{AC}}, \bm{p}_i^{\text{F}}, \bm{p}_i^{\text{CH}}, \bm{p}_i^{\text{DIS}}, \bm{e}_i^{\text{B}}, \bm{p}_i^{\text{ET}}, \forall i \in \mathcal{N} \}.
        \end{aligned} 
    \end{equation*}
    
Therefore, solving \textbf{TCMP} is to determine the optimal energy scheduling $ \{ \bm{p}_i^{\text{RE},\ast}, \bm{p}_i^{\text{G},\ast}, \bm{p}_i^{\text{AC},\ast}, \bm{p}_i^{\text{F},\ast}, \bm{p}_i^{\text{CH},\ast}, \bm{p}_i^{\text{DIS},\ast}, \bm{e}_i^{\text{B},\ast} \}$ and the optimal energy trading $\{ \bm{p}_i^{\text{ET},\ast} \}$ for each user $i \in \mathcal{N}$. Let $C_i^{\text{O},\ast} {\triangleq} C_i^{\text{O}}(\bm{p}_i^{\text{G},\ast}, \bm{p}_i^{\text{AC},\ast}, \bm{p}_i^{\text{F},\ast}, \bm{p}_i^{\text{CH},\ast}, \bm{p}_i^{\text{DIS},\ast})$ denote the optimized operating cost of user $i$ in the energy-trading scenario. Substituting the optimal energy scheduling and trading decisions into Problem \textbf{ETOP} leads to Problem \textbf{TBAP} as follows. 
    
\textbf{TBAP}: Trading Benefit Allocation Problem
    \begin{equation*}
        \begin{aligned}
            &\text{maximize} &&
            \prod\nolimits_{i\in\mathcal{N}} \left[ \Delta_i - C_{i}^{\text{ET}} (\boldsymbol{\pi}_i^{\text{ET}}) \right]  \\
            &\text{subject to} &&
            \text{\eqref{constraint-trading2}},~\text{\eqref{constraint-trading3}}\\
            &\text{variables:} &&
            \{ \boldsymbol{\pi}_i^{\text{ET}},~ \forall i \in \mathcal{N} \},
        \end{aligned}
    \end{equation*}
where $\Delta_i \triangleq \bar{C}_i^{\text{O}} - C_i^{\text{O},\ast}$ denotes the difference between the minimized operating cost of user $i$ in Problem \textbf{EMP$_i$} and Problem \textbf{TCMP}.

\subsection{Decentralized Algorithm Design}\label{subsec:algorithm}

To solve subproblems \textbf{TCMP} and \textbf{TBAP} on the blockchain-enabled transactive energy platform, we design a decentralized algorithm that suits the operation of the blockchain. In this algorithm, the users locally optimize their own energy usage and communicate with the blockchain on their energy schedule and payment decisions. By doing so, users do not reveal their private information regarding their operational parameters during the trading process. For practical implementation on the blockchain, our decentralized algorithm consists of two levels: 1) the lower-level problem allows users to optimize their trading and payment decisions; 2) the higher-level problem collects the decisions from users to achieve trading consensus.
    
First, we employ the alternating direction method of multipliers (ADMM) method \cite{boyd2011distributed} to solve \textbf{TCMP}, as ADMM has a good convergence for convex optimization problems with non-strictly convex objective functions. We introduce auxiliary variables $\hat{\boldsymbol{p}}_i^{\text{ET}} = \{\hat{p}_{i,j}^{\text{ET}}[t], \forall j \in \mathcal{N} \backslash i, \forall t \in \mathcal{H}\}$ to denote the energy trading decisions, and rewrite constraints \eqref{constraint-trading1} as
    \begin{align}
        \hat{p}_{i,j}^{\text{ET}}[t] =\ & p_{i,j}^{\text{ET}}[t],\ \forall j \in \mathcal{N} \backslash i,~\forall i \in \mathcal{N},~\forall t \in \mathcal{H}, \label{constraint-auxiliary1}\\
        \hat{p}_{i,j}^{\text{ET}}[t] + \hat{p}_{j,i}^{\text{ET}}[t] =\ & 0,\ \forall j \in \mathcal{N} \backslash i,~\forall i \in \mathcal{N},~\forall t \in \mathcal{H}. \label{constraint-auxiliary2}
    \end{align}
The higher-level algorithm discussed later handles the update of auxiliary variables $\hat{p}_{i,j}^{\text{ET}}[t]$ and enforces the energy trading clearing constraint in \eqref{constraint-auxiliary2}.
    
To decompose the coupling constraint of the auxiliary variables $\hat{p}_{i,j}^{\text{ET}}[t]$ and the real trading decisions $p_{i,j}^{\text{ET}}[t]$ in \eqref{constraint-auxiliary1}, we introduce dual variables $\boldsymbol{\lambda} = \{ \bm{\lambda}_{i},~\forall i \in \mathcal{N} \}$ where $\bm{\lambda}_{i} = \{ \lambda_{i,j}^{t},~\forall j \in \mathcal{N} \backslash i,~t \in \mathcal{H}\}$ for constraints \eqref{constraint-auxiliary1}. We then obtain the augmented Lagrangian for Problem \textbf{TCMP} as
    \begin{align*}
        &L^{\text{TCMP}} =  \sum\nolimits_{i\in\mathcal{N}} 
        C_i^{\text{O}}(\bm{p}_i^{\text{G}}, \bm{p}_i^{\text{AC}}, \bm{p}_i^{\text{F}}, \bm{p}_i^{\text{CH}}, \bm{p}_i^{\text{DIS}}) \\
        &+ {\sum_{i\in\mathcal{N}}} {\sum_{j \in \mathcal{N} \backslash i}} \sum_{t\in\mathcal{H}} 
        \left[ \frac{\rho_{1}}{2} \left( \hat{p}_{i,j}^{\text{ET}}[t] {-} p_{i,j}^{\text{ET}}[t] \right)^{2} 
        {+} \lambda_{i,j}^{t} \left( \hat{p}_{i,j}^{\text{ET}}[t] {-} p_{i,j}^{\text{ET}}[t] \right) 
        \right],
    \end{align*}
where $\rho_{1} >0$ is a coefficient for the quadratic penalty of \eqref{constraint-auxiliary1}.
    
To solve Problem \textbf{TCMP}, we decompose it into a lower-level problem and a higher-level problem. In the lower-level problem, users individually optimize their cost in parallel given dual variables $\boldsymbol{\lambda}$ and auxiliary variables $\boldsymbol{\hat{p}}_i^{\text{ET}},~\forall i \in \mathcal{N}$. The higher-level problem updates the auxiliary variables and dual variables based on the trading decisions of users. The variable exchange between these two levels is done by sending transactions via the blockchain.
    
Specifically, in the lower-level problem, given the dual variables $\lambda_{i,j}^{t}$ and auxiliary variables $\hat{p}_{i,j}^{\text{ET}}[t], j {\in} \mathcal{N} \backslash i$, user $i$ solves the following optimization problem:
    
\textbf{LLP1$_i$}: Lower-level problem of TCMP
    \begin{equation*}
        \begin{aligned}
            &\text{minimize} && 
            C_i^{\text{O}}(\bm{p}_i^{\text{G}}, \bm{p}_i^{\text{AC}}, \bm{p}_i^{\text{F}}, \bm{p}_i^{\text{CH}}, \bm{p}_i^{\text{DIS}}) \\
            &&& + \sum_{j \in \mathcal{N} \backslash i} \sum_{t\in\mathcal{H}} 
        \left[ \frac{\rho_{1}}{2} \left( \hat{p}_{i,j}^{\text{ET}}[t] - p_{i,j}^{\text{ET}}[t] \right)^{2} 
        - \lambda_{i,j}^{t}  p_{i,j}^{\text{ET}}[t] \right] \\
            &\text{subject to} &&
            \text{\eqref{constraint-load1}},~ \text{\eqref{constraint-load2} to \eqref{constraint-load4}},~\text{\eqref{constraint-load6} to \eqref{constraint-load5}},~ \text{\eqref{constraint-load7} to \eqref{constraint-load10}},~\text{\eqref{constraint-load12}}\\
            &\text{variables:} &&
            \bm{p}_i^{\text{RE}}, \bm{p}_i^{\text{G}}, \bm{p}_i^{\text{AC}}, \bm{p}_i^{\text{F}}, \bm{p}_i^{\text{CH}}, \bm{p}_i^{\text{DIS}}, \bm{e}_i^{\text{B}}, \bm{p}_i^{\text{ET}}.
        \end{aligned} 
    \end{equation*}
By solving \textbf{LLP1}, user $i$ obtains the optimal energy schedule including the energy trading decision $\bm{p}_i^{\text{ET}}$. The user then sends $\bm{p}_i^{\text{ET}}$ to the higher-level problem for further interaction.
    
On the other hand, the higher-level problem solves the dual variables $\boldsymbol{\lambda}$ and auxiliary variables $\boldsymbol{\hat{p}}_i^{\text{ET}}$ for all users $i \in \mathcal{N}$. The higher-level problem of \textbf{TCMP} is formulated as follows:
    
    \textbf{HLP1}: Higher-level problem of TCMP
    \begin{equation*}
        \begin{aligned}
        &\text{minimize} && \sum_{i\in\mathcal{N}} \sum_{j \in \mathcal{N} \backslash i} \sum_{t\in\mathcal{H}} 
        \left[ \frac{\rho_{1}}{2} \left( \hat{p}_{i,j}^{\text{ET}}[t] - p_{i,j}^{\text{ET}}[t] \right)^{2} 
        + \lambda_{i,j}^{t} \hat{p}_{i,j}^{\text{ET}}[t] \right] \\
        &\text{subject to} &&
        \text{\eqref{constraint-auxiliary2}} \\
        &\text{variables:} &&
        \{ \hat{\bm{p}}_i^{\text{ET}},~ \forall i \in \mathcal{N} \}.
        \end{aligned} 
    \end{equation*}
    
Solving \textbf{HLP1} outputs the optimal auxiliary variables
        \begin{align}
            \begin{split}
                \hat{p}_{i,j}^{\text{ET}}[t] {=} -\hat{p}_{j,i}^{\text{ET}}[t]
                {=} \frac{\rho_{1} \left( p_{i,j}^{\text{ET}}[t] - p_{j,i}^{\text{ET}}[t] \right) - \left( \lambda_{i,j}^{t} - \lambda_{j,i}^{t} \right) }{2 \rho_{1}}, \label{updateenergy}
            \end{split}
        \end{align}
and updates the dual variables as follows
    \begin{align}
        \lambda_{i,j}^{t} \leftarrow \lambda_{i,j}^{t} + \rho_{1} \left( \hat{p}_{i,j}^{\text{ET}}[t] - p_{i,j}^{\text{ET}}[t] \right). \label{updatelambda}
    \end{align}

Similarly, we can derive the lower-level problem and higher-level problem for Problem \textbf{TBAP}. We introduce auxiliary variables $\boldsymbol{\hat{\pi}^{\text{ET}}} {=} \{\hat{\pi}_{i,j}^{\text{ET}},\forall i {\in} \mathcal{N},j {\in} \mathcal{N} \backslash i\}$ and rewrite the trading payment clearing constraints \eqref{constraint-trading2} as
    \begin{align}
        & \hat{\pi}_{i,j}^{\text{ET}} = \pi_{i,j}^{\text{ET}},~\forall j \in \mathcal{N} \backslash i,~\forall i \in \mathcal{N}, \label{constraint-auxiliary3} \\
        & \hat{\pi}_{i,j}^{\text{ET}} + \hat{\pi}_{j,i}^{\text{ET}} = 0,~\forall j \in \mathcal{N} \backslash i,~\forall i \in \mathcal{N}, \label{constraint-auxiliary4} 
    \end{align}
and we introduce $\gamma_{i,j}$ as dual variables for constraints \eqref{constraint-auxiliary3}.

Due to the space limitation, we omit the intermediate steps and only present the results as follows. In the lower-level problem, given $\gamma_{i,j}$ and $\{ \hat{\pi}_{i,j}^{\text{ET}},~j \in \mathcal{N} \backslash i \}$, user $i$ solves the following optimization problem: 

\textbf{LLP2$_i$}: Lower-level problem of TBAP
    \begin{equation*}
        \begin{aligned}
            & \text{minimize}
            && - \ln \left[ \Delta_i - C_{i}^{\text{ET}} (\boldsymbol{\pi}_i^{\text{ET}}) \right] \\
            &&& + \sum\nolimits_{j \in \mathcal{N} \backslash i}  
            \left[   
            \frac{\rho_{2}}{2} \left( \hat{\pi}_{i,j}^{\text{ET}} - \pi_{i,j}^{\text{ET}}  \right)^{2} 
            - \gamma_{i,j} \pi_{i,j}^{\text{ET}}
            \right]  \\
            & \text{subject to} 
            && \text{\eqref{constraint-trading3}}\\
            & \text{variables:} 
            && \boldsymbol{\pi}_{i}^{\text{ET}},
        \end{aligned}
    \end{equation*}
where $\rho_{2} >0$ is a penalty coefficient.

The higher-level problem \textbf{HLP2} updates $\hat{\pi}_{i,j}^{\text{ET}}$ and $\gamma_{i,j}$ as
    \begin{align}
            \begin{split}
                & \hat{\pi}_{i,j}^{\text{ET}} = -\hat{\pi}_{j,i}^{\text{ET}}
                =  \frac{\rho_{2} \left( \pi_{i,j}^{\text{ET}} - \pi_{j,i}^{\text{ET}} \right) - \left( \gamma_{i,j} - \gamma_{j,i} \right) }{2 \rho_{2}}, \label{updatepayment}
            \end{split}
        \end{align}
    \begin{equation}
        \gamma_{i,j} \leftarrow \gamma_{i,j} + \rho_{2} 
        \left( \hat{\pi}_{i,j}^{\text{ET}} - \pi_{i,j}^{\text{ET}} \right), \label{updategamma}
    \end{equation}
when receiving users' payment decisions $\pi_{i,j}^{\text{ET}}$.

\begin{figure}[!t]
    \centering
    \includegraphics[width=8.6cm]{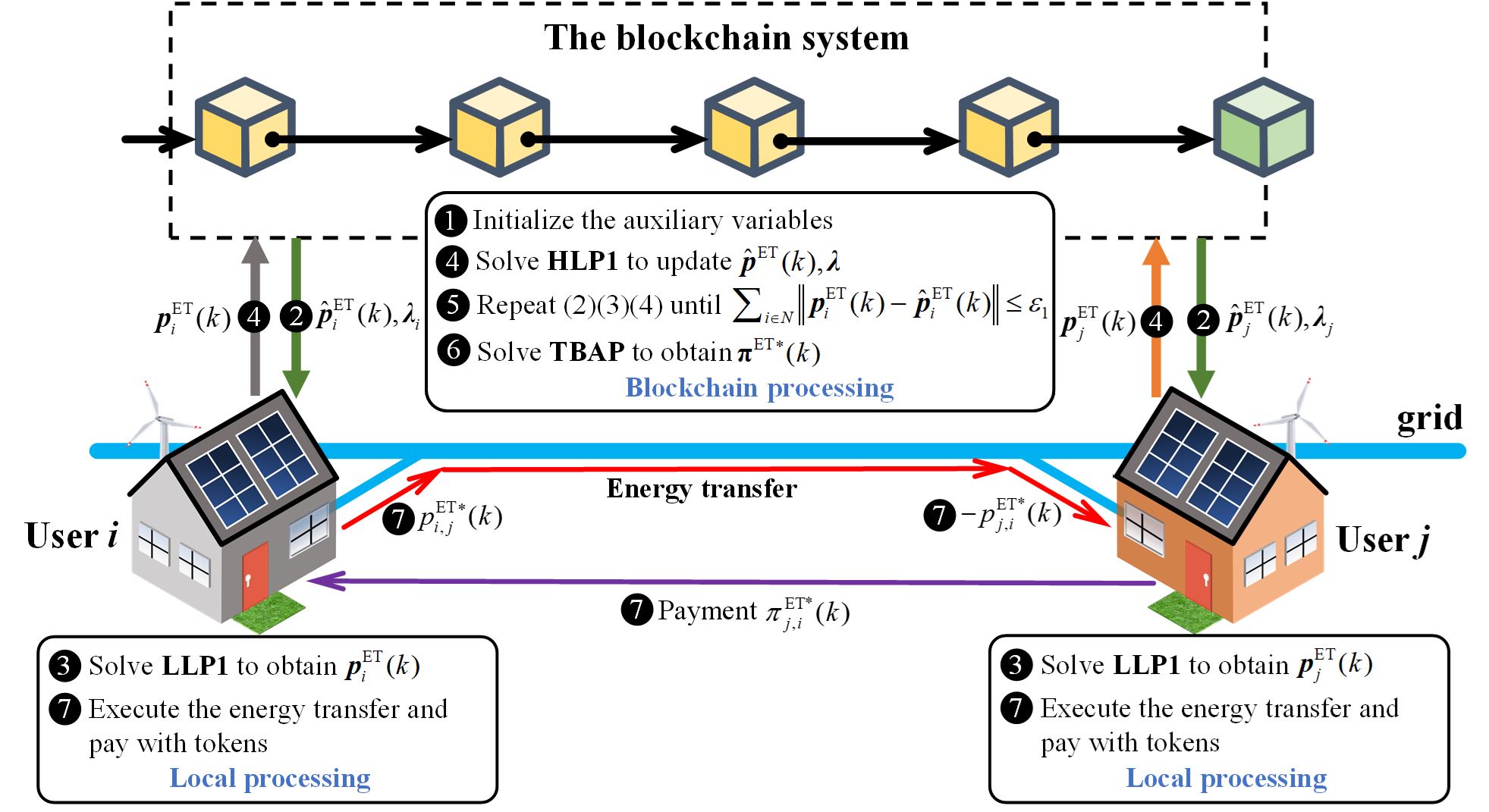}
    \caption{The energy trading process on the blockchain-empowered transactive energy platform.}
    \vspace*{-3mm}
    \label{f4:process}
\end{figure}
To sum up, we illustrate the process of the decentralized P2P energy trading algorithm in Fig.~\ref{f4:process} and Algorithm~\ref{alg1}. Specifically, the algorithm solves the higher-level and lower-level problems for both \textbf{TCMP} and \textbf{TBAP} in an iterative fashion. As shown in Fig.~\ref{f4:process}, in Step~\textcircled{\small{1}}, the smart contract initializes all the auxiliary variables at the beginning of the day, and then notices the auxiliary variables $\bm{\hat{p}}_i^{\mathrm{ET}(k)}$ and $\bm{\lambda}_i$ to user $i,~\forall i {\in} \mathcal{N}$ in Step~\textcircled{\small{2}}. Upon receiving the auxiliary variables, each user $i,~\forall i {\in} \mathcal{N}$ computes its trading demand $\bm{p}_{i}^{\text{ET}}(k)$ by solving the Problems \textbf{LLP1$_i$} in Step~\textcircled{\small{3}} locally, and then reports $\bm{p}_{i}^{\text{ET}}(k)$ to the smart contract in Step~\textcircled{\small{4}}. Then the smart contract solves the higher-level problem \textbf{HLP1} with $\bm{p}_{i}^{\text{ET}}(k)$, and updates the market information including auxiliary variables and dual variables in \eqref{updateenergy}, \eqref{updatelambda}, \eqref{updatepayment}, and \eqref{updategamma}, then broadcasts the updated market information to all the users for the next iteration. Since Algorithm~\ref{alg1} runs in an iterative manner, Step \textcircled{\small{2}},\textcircled{\small{3}},\textcircled{\small{4}} are repeated until Problem \textbf{TCMP} converges. When Problem \textbf{TCMP} converges, the optimal energy trading scheduling $\bm{p}_i^{\text{ET}}$ is obtain for every user $i,~\forall i {\in} \mathcal{N}$ in Step~\textcircled{\small{5}}. Next, the Problem \textbf{TBAP} is also solved in a similar iterative method in Step~\textcircled{\small{6}} and the optimal energy trading payments $\bm{\pi}_i^{\mathrm{ET}}$ for all the users are obtained. In Step \textcircled{\small{7}}, each user executes its energy trading schedule $\bm{p}_i^{\text{ET}},~\forall i {\in} \mathcal{N}$, and pays to or receives payments from other users at the end of the day.

During the iteration of the decentralized algorithm, the users do not need to reveal the process of optimizing the trading decisions, so that their privacy is well preserved. According to \cite{boyd2011distributed}, the two-level decentralized algorithm converges to the optimal solution of Problems \textbf{TCMP} and \textbf{TBAP} if we choose the stepsizes $\rho_{1}(k) = 1/k$ and $\rho_{2}(k) = 1/k$, in which $k$ denotes the number of iterations. The result of this algorithm optimizes both the individual user's benefit and the overall welfare of the system.
    
\begin{algorithm}[!t]
     \caption{Decentralized energy trading algorithm}
     \label{alg1} 
     \SetAlgoLined
     (1) Solve Problem \textbf{TCMP}\\     
     \textbf{Initialization}: $k {\leftarrow} 1$, $\rho_{1}(0) {\leftarrow} 1$, $\boldsymbol{\lambda}(0) {\leftarrow} \textbf{0}$;
    
    \While{$\sum_{i \in \mathcal{N}} \parallel \bm{\hat{p}}_i^{\mathrm{ET}}(k) - \bm{p}_i^{\mathrm{ET}}(k) \parallel > \epsilon_{1}$}{
    \hspace*{-1.1em}$\neg$ \For{$i \in \mathcal{N}$}{
        \hspace*{-1.1em}$\neg$ User $i$ solves \textbf{LLP1$_i$} based on $\hat{\bm{p}}_i^{\text{ET}}(k{-}1)$, $\bm{\lambda}_i(k{-}1)$;
        
        \hspace*{-1.1em}$\neg$ User $i$ updates $\bm{p}_i^{\text{ET}}(k)$ to \textbf{HLP1};
    }
    
    \hspace*{-1.1em}$\neg$ \textbf{HLP1} updates $\bm{\hat{p}}_{i}^{\text{ET}}(k),\forall i \in \mathcal{N}$ and $\bm{\lambda}(k)$;
    
    \hspace*{-1.1em}$\neg$ $k \leftarrow k+1$;
    }
    
    (2) Solve Problem \textbf{TBAP} \\
     \textbf{Initialization}:  $k {\leftarrow} 1$,$\rho_{2}(0) {\leftarrow} 1$, $\boldsymbol{\gamma}(0) {\leftarrow} \textbf{0}$;
    
    \While{$\sum_{i \in \mathcal{N}} \parallel \bm{\hat{\pi}}_i^{\mathrm{ET}}(k) - \bm{\pi}_i^{\mathrm{ET}}(k) \parallel > \epsilon_{2}$}{
    
    \hspace*{-1.1em}$\neg$ \For{$i \in \mathcal{N}$}{
    
        \hspace*{-1.1em}$\neg$ User $i$ solves \textbf{LLP2$_i$} based on $\bm{\hat{\pi}}_i^{\text{ET}}(k)$, $\bm{\gamma}_i(k)$;
        
        \hspace*{-1.1em}$\neg$ User $i$ updates $\bm{\pi}_i^{\text{ET}}(k)$ to \textbf{HLP2};
    }
    
    \hspace*{-1.1em}$\neg$ \textbf{HLP2} updates $\bm{\hat{\pi}}_{i}^{\text{ET}},\forall i \in \mathcal{N}$ and $\bm{\gamma}$;
    
    \hspace*{-1.1em}$\neg$ $k \leftarrow k+1$;
    }
 \KwResult{$\bm{p}_i^{\text{ET},\ast}$ and $\bm{\pi}_i^{\text{ET},\ast}$, $\forall i \in \mathcal{N}$.} 
\end{algorithm}

\section{System Implementation and Evaluation}\label{sec:eval}

We implement the decentralized energy trading algorithm (namely Algorithm~\ref{alg1}) on the blockchain-empowered transactive energy platform, as shown in Fig.~\ref{f4:process}. The smart contract deployed on the blockchain handles \textbf{HLP1} and \textbf{HLP2}, while users handle \textbf{LLP1$_i$} and \textbf{LLP2$_i$} locally. When the algorithm converges, the users execute their energy trading decisions and pay with the blockchain tokens at the end of the trading day.
\vspace*{-2mm}

\subsection{System Design and Implementation}\label{sec:blockchain}

\begin{figure}[!t]
    \centering
    \includegraphics[width=8.7cm]{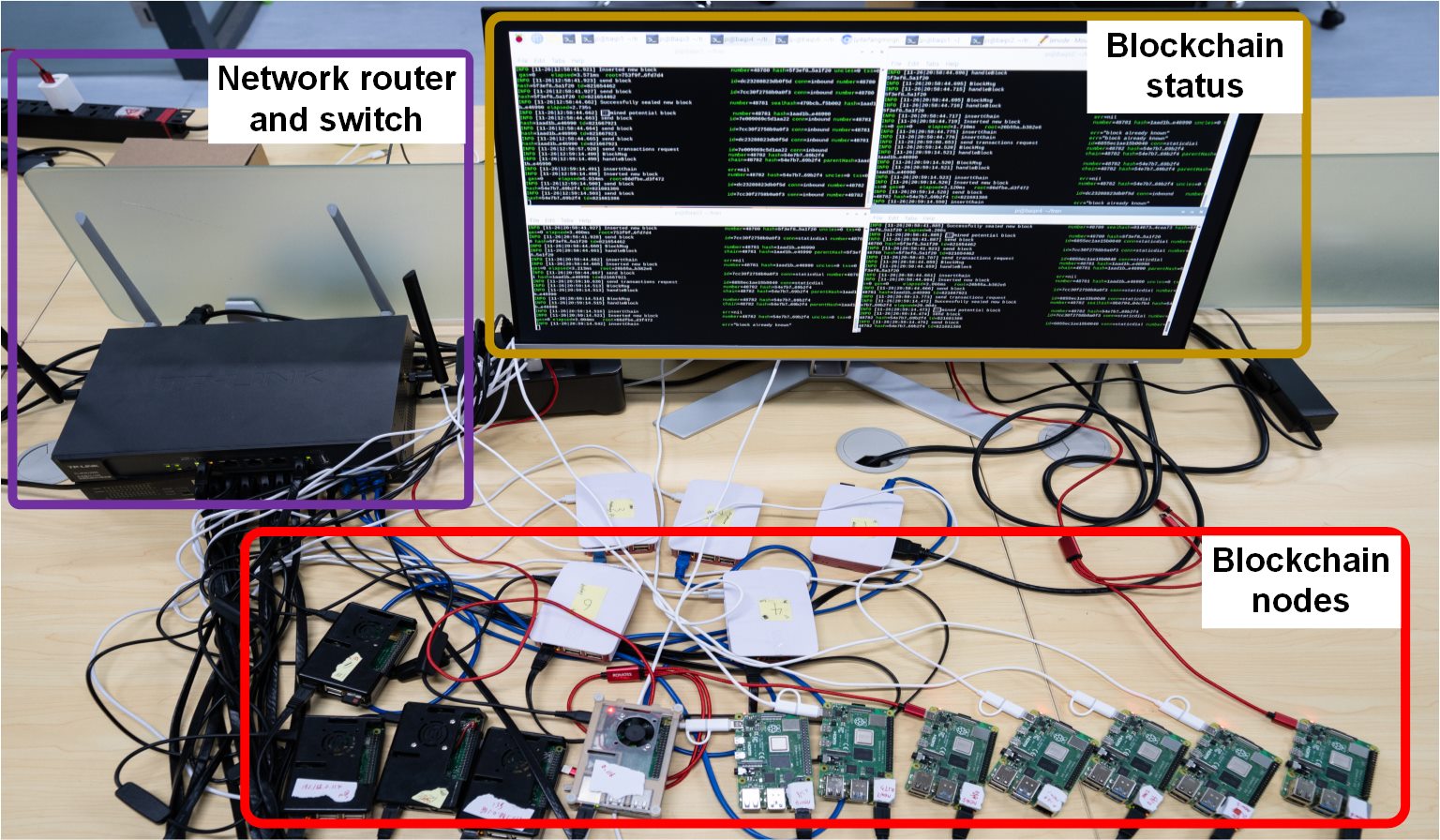}
    \vspace*{-2mm}
    \caption{The prototype blockchain system built on IoT devices.}
    \vspace*{-3mm}
    \label{f4:demo}
\end{figure}

To verify our design of the energy trading platform and evaluate the performance of the blockchain network, we develop a prototype blockchain system on a small-scale network with 18 IoT nodes, as shown in Fig.~\ref{f4:demo}. We let five nodes be the validators (white) and let the rest (black) be the normal users. Each node is a Raspberry~Pi Model~3B+ \cite{rpi} with a quadcore ARM Cortex-A53 CPU, a 16GB TF card, and 1GB DDR2 SDRAM. 

We modify the source code of the Quorum \cite{quorum} project to support energy trading transactions and token transfer for payments. We choose Quorum as our blockchain platform based on the following considerations. First, the performance of Quorum is better than other available blockchains such as IOTA and Ethereum. Second, Quorum is more suitable on IoT devices because Quorum's PBFT consensus protocol consumes less memory and CPU than PoW-based blockchains. 
Third, Quorum supports the smart contract to implement the proposed decentralized energy trading algorithm.

In our prototype system, a validator node consumes 480MB memory, and a client node consumes less than 200MB memory. The transaction transmission time is about 5ms during our test, and each block can contain 2000 transactions at most. The consensus time is between 10ms and 100ms, and the measured TPS (transaction per second) is around 700 during our test. To evaluate the decentralized optimization algorithm that solves {TCMP} and {TBAP}, we implement Algorithm~\ref{alg1} in Matlab and upload the results to the smart contract in each iteration. 

\vspace*{-2mm}
\subsection{Performance Evaluation}

\subsubsection{Simulation Setup}
We evaluate the decentralized optimization algorithm with real data of renewable generation \cite{wang2017joint} and household load in the city of Austin, Texas by Pecan Street \cite{pecan} and . The data include power consumption, renewable energy generation (solar and wind), and outdoor temperature from September 6 to September 12 in 2016. We randomly select 10 users from the dataset and feed their data into the system described in Section~\ref{sec:blockchain}. 
Due to the space limit, we list the parameters used in our simulation in the supplementary material.

\begin{figure}[!t]
    \centering
    \includegraphics[width=8.5cm]{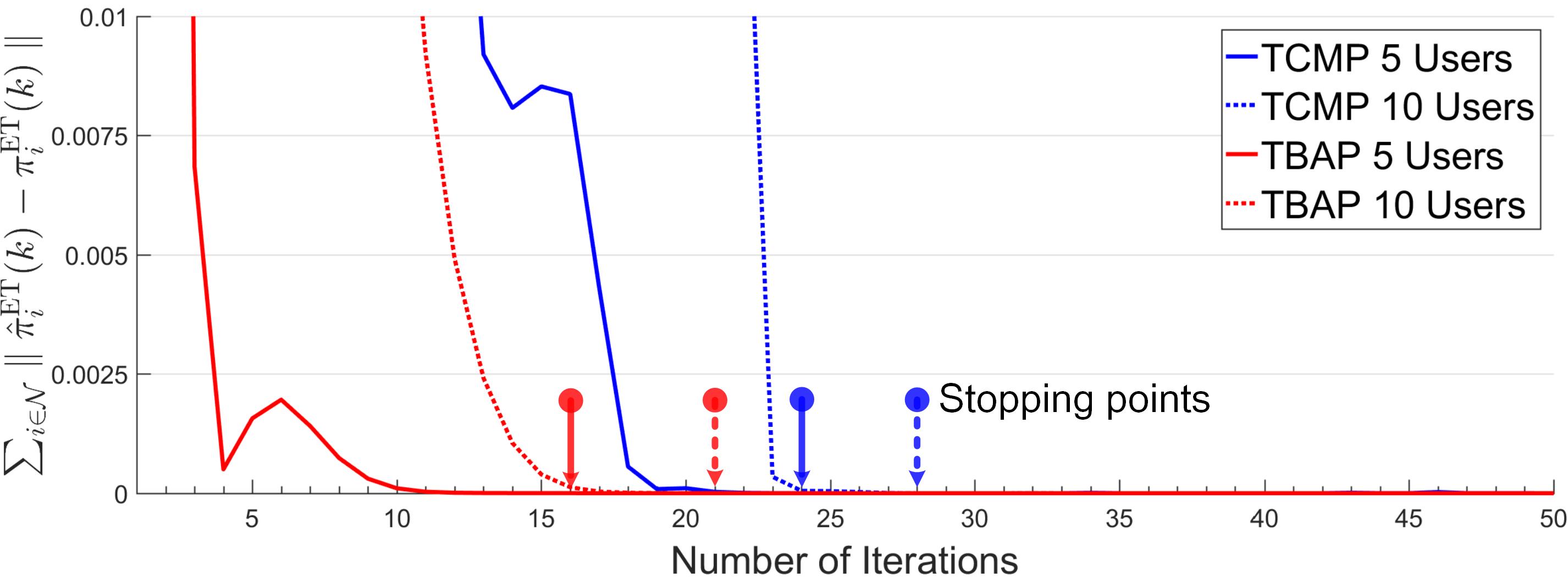}
    \caption{The convergence speed of the decentralized energy trading algorithm. Here we simulate with both five users and ten users.}
    \label{f:converge}
\end{figure}

\subsubsection{Algorithm Convergence}
We test the convergence speed of Algorithm~\ref{alg1} with 5 users and 10 users, respectively. We set the convergence thresholds of TCMP ($\epsilon_1$) and TBAP ($\epsilon_2$) to $1 \times 10^{-6}$. As shown in Fig.~\ref{f:converge}, the errors of algorithm TCMP and TBAP are plotted with respect to the number of iterations. The TCMP and TBAP algorithms converge at iteration 24 and 16 in the case of 5 users, and at iteration 28 and 21 in the case of 10 users. \rv{To measure the time consumed by the two algorithms in the case of 10 users, we test the lower-level algorithm (LLP1 and LLP2) on the Raspberry Pi node, and the higher-level algorithm (HLP1 and HLP2) on a PC with Intel i7-8700 CPU. The results show that the lower-level algorithms take 2s for each iteration, and the higher-level algorithms take 25s for each iteration.} The simulation results show that the proposed decentralized optimization algorithm converges fast in two cases with real-world data, which is of practical importance.

\begin{figure}[!t]
    \centering
    \includegraphics[width=8.7cm]{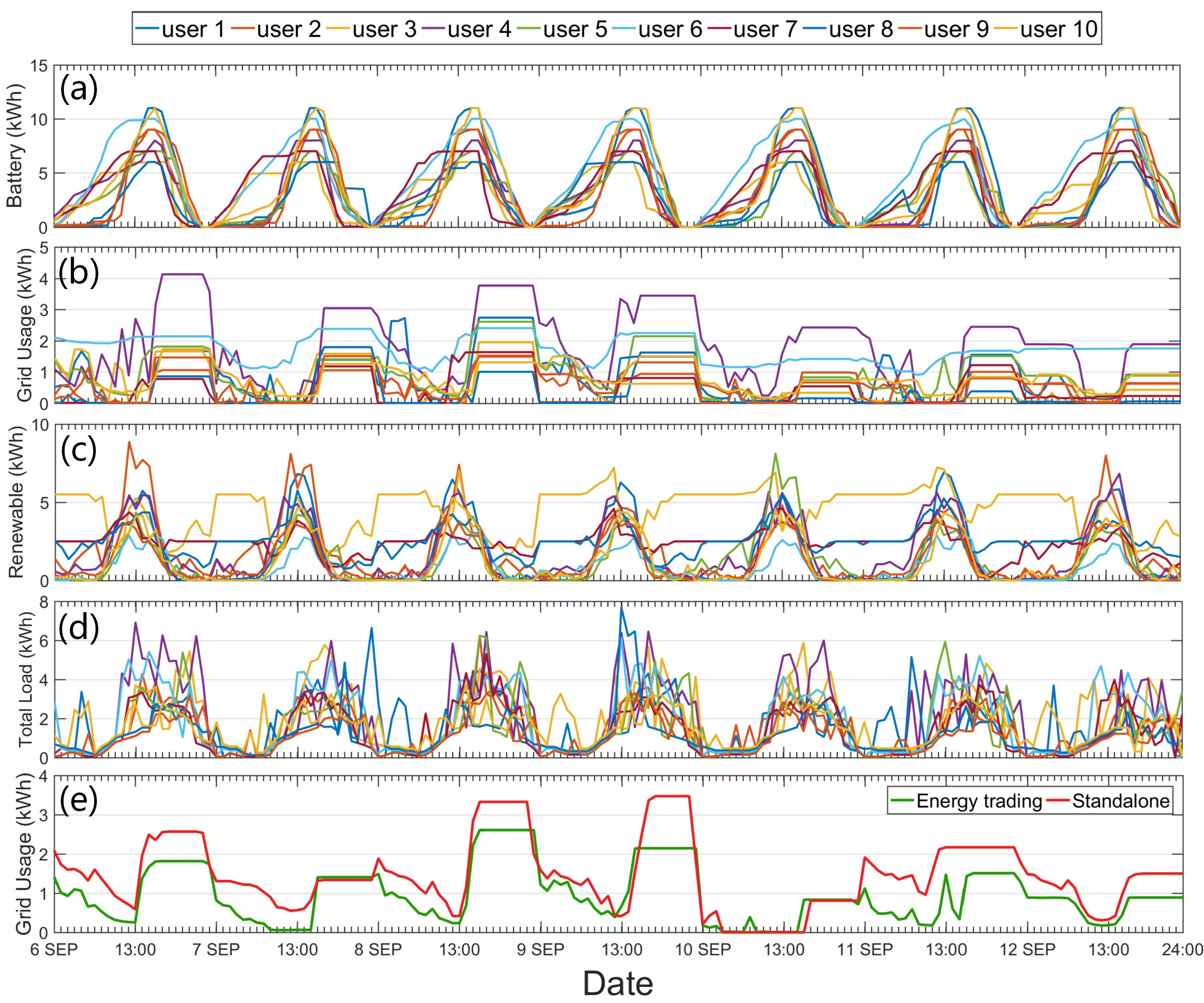}   
    \vspace*{-3mm}
    \caption{The power scheduling of the users for (a) battery energy level, (b) grid power usage, (c) renewable energy supply, (d) total load, (e) user 5's grid power consumption in both scenarios.}
    \label{f:usage}
    \vspace*{-4mm}
\end{figure}

\subsubsection{Power Scheduling in the Energy-Trading Scenario} Fig.~\ref{f:usage} shows the optimized decisions of all the 10 users over one week (September 6-12, 2017) solved by Algorithm~\ref{alg1}. We see that users tend to store excess renewable energy into their batteries in the daytime and discharge the batteries to meet the demand in the evening and at night. Fig.~\ref{f:usage}e compares the energy purchased from the gird for user 5 in the two scenarios as an example. It shows that the user's grid power consumption is effectively reduced in the energy-trading scenario.

\begin{figure}[!t]
    \centering
    \includegraphics[width=8.8cm]{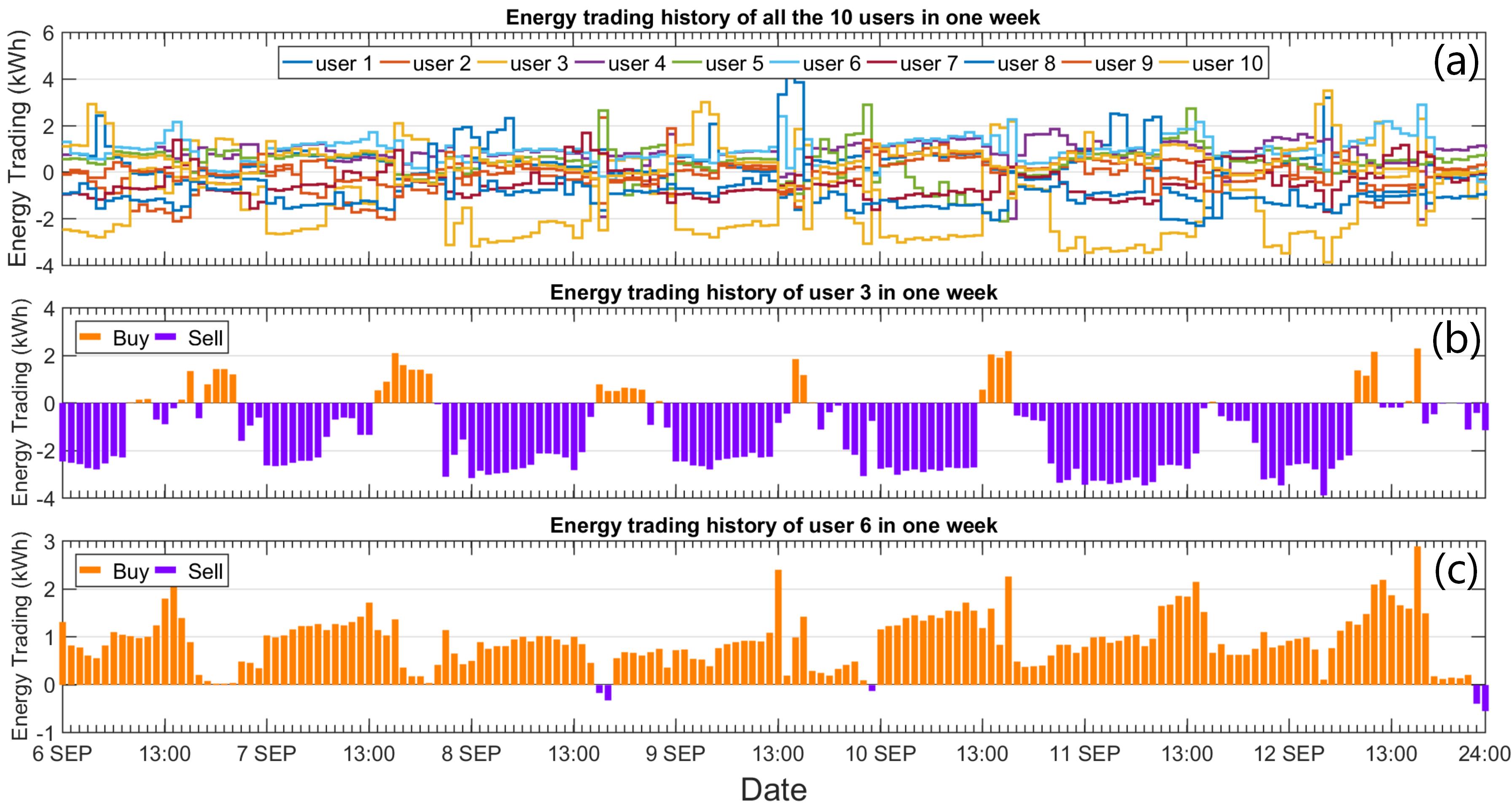}
    \caption{The energy trading results: (a) total energy traded per hour for all users (positive value denotes buying, negative value denotes selling). (b) and (c) are the energy trading results for user 3 and user 6, respectively.}
    \label{f:trading}
\end{figure}

\subsubsection{Energy Trading Performance} We plot the energy trading decisions of all the 10 users with one-hour granularity in Fig.~\ref{f:trading}a. The figure shows active energy trading activities among users throughout the whole week. We further show the energy trading decisions of two typical users (user 3 and user 6) in Fig.~\ref{f:trading}b and Fig.~\ref{f:trading}c, respectively. We see complementary patterns of these two users though they can trade with all other users not limited to these two users. User 6 is short of energy in the daytime and thus purchases energy from other users frequently. By contrast, user 3 is at most of the time self-sustained and sells a lot of generated energy to other users. Our blockchain-empowered trading system provides an effective platform for users to interact with each other and exchange energy for mutual benefits.

\begin{figure}[!t]
    \centering
    \includegraphics[width=8.7cm]{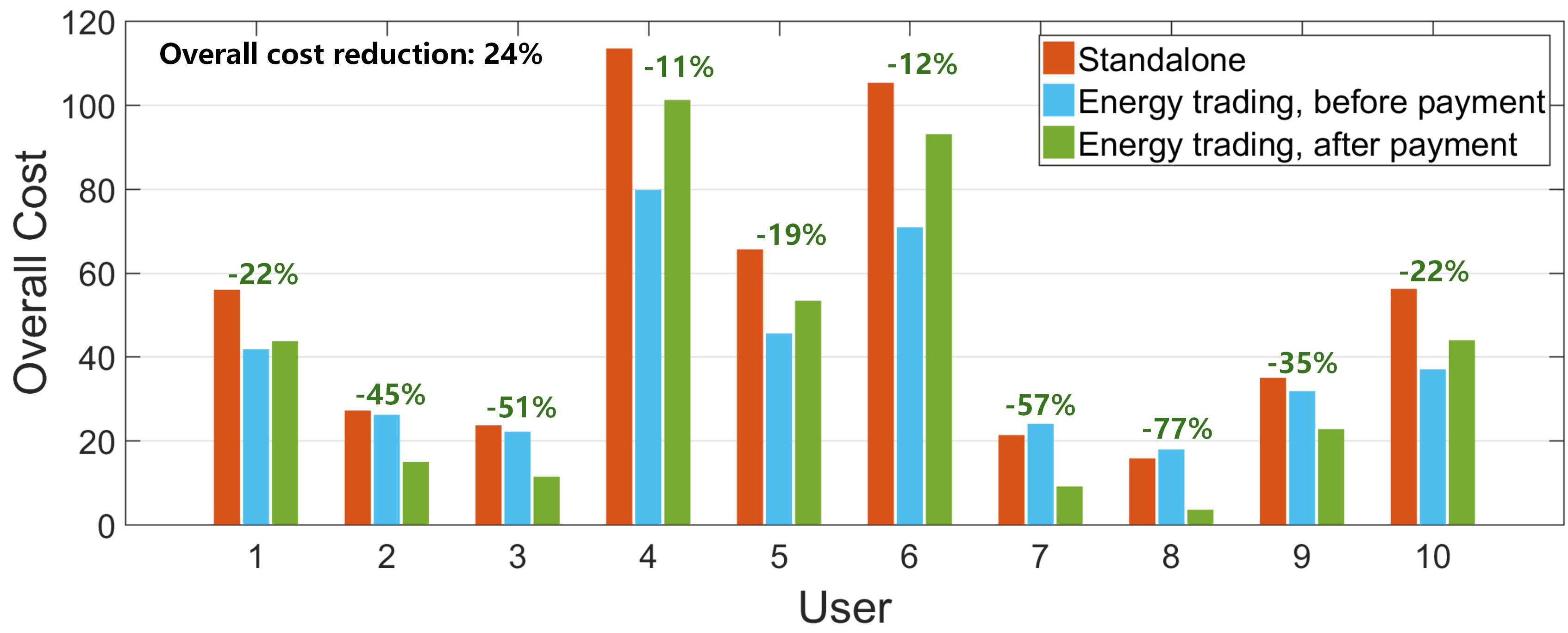}
    \vspace*{-3mm}
    \caption{The total costs of the ten users in one week. We compare the costs of standalone scenario (orange), energy-trading scenario before payment (purple), and energy-trading scenario after payment (green).}
    \vspace*{-4mm}
    \label{f:overall_cost}
\end{figure}

\subsubsection{Cost Reduction}
To evaluate the benefits that the users can obtain by joining the energy trading platform, we compare each user's total cost with and without energy trading in Fig.~\ref{f:overall_cost}. The orange bars show the users' costs in the standalone scenario (without energy trading) optimized in the EMP problem (described in Section~\ref{sec:standalone}); the purple bars show the users' costs in the energy-trading scenario before payment according to TCMP problem; the green bars show the costs with energy trading after payment solved in the TBAP problem. Comparing the orange bars with purple bars, we see that most of the users reduce their operating costs directly by energy trading. Note that user 7's and user 8's costs increase because they both sell much energy to others, and their loss will be compensated by receiving payments from other users. We annotate the reduction of users' costs in the energy-trading scenario after payment with respect to the standalone scenario in green above the bars. It is shown that all users' costs are reduced after payment, and the overall cost of the system is reduced by 24\% in our simulation. This result shows that the proposed energy trading algorithm effectively reduces the users' cost, and can well incentivize users to join our transactive energy platform.

\section{Conclusion}\label{sec:conclusion}
This paper presented a blockchain-empowered transactive energy platform that allows prosumers to trade electricity with each other via a blockchain network. We employed blockchain, an emerging technology in cryptocurrency, to make the transactive energy platform a decentralized system that is robust and trustable. Furthermore, we designed a decentralized energy trading algorithm that well preserves the users' privacy. The energy trading algorithm achieves the socially optimal solution and also incentivizes prosumers by reducing the cost of every participant. We tested the transactive energy platform on a prototype blockchain network with 18 nodes. The results showed that our blockchain-based transactive energy platform is feasible on practical IoT devices (e.g., smart meters). Simulations using real-world data showed that users' individual costs are reduced by up to 77\%, and the overall cost of the platform is reduced by 24\%. 

In our future work, we will study more efficient decentralized optimization algorithms that can scale up to hundreds or even thousands of users. 
We will explore methods that can fully implement complex algorithms in smart contracts.


\bibliographystyle{IEEEtran}
\bibliography{bibfile} 

\end{document}